# Segmented Continuous Optimization: A novel framework for continuous non-linear curve fitting

Teymur Aghayev
Dept. of Electronic Systems
Vilnius Gediminas Technical University
Vilnius, Lithuania
https://orcid.org/0009-0007-6007-2785

***Abstract*—Segmented curve fitting remains an essential approach for the comprehensive analysis of local patterns in non-stationary time-series data. However, traditional regression algorithms primarily focus on linear or polynomial functions, which can be insufficient for analyzing raw signals with oscillatory or transcendental behavior. In this paper, we propose Segmented Continuous Optimization (SCO), a framework that performs piecewise continuous curve fitting on various non-linear models, including trigonometric, polynomial, and exponential. SCO presents a novel signal representation by optimizing a user-defined model in segments with $C^1$ continuity to properly analyze the data's local and global trends. The framework is tested for accuracy and efficiency across all included models. Finally, we provide examples using velocity and EEG datasets to demonstrate the algorithm's practical usage in examining signal patterns, optimized parameters, derivatives, and integrals of the final fit.**

## I. Introduction

Many real-world signals are modeled using piecewise analytical functions. Such a segmented representation has found broad applications in the investigation of signals across molecular biology, medicine, climatology, engineering, and econometrics [1,2,3]. Various algorithms have been proposed, including Muggeo's segmented regression, Multivariate Adaptive Regression Splines (MARS) for linear modeling [4,5], and High-Order Polynomial Segmenter (HOPS), trend filtering, cubic splines for polynomial modeling [6,7,8]. However, traditional regression algorithms mainly focus on polynomial or linear functions [9], which can be insufficient for examining real-world signals that often exhibit oscillatory or transcendental behavior. As a result, developing a generalized framework that ensures continuity while accounting for user-defined models is essential to address current limitations. We introduce Segmented Continuous Optimization (SCO), a novel framework that performs segmented curve fitting with $C^1$ continuity to provide a more relevant, analytical representation of complex signals. The framework executes a Levenberg-Marquardt [10,11] optimization across segments for solving a non-linear least squares problem; continuity is enforced by algebraically deriving parameters during the calculation of residuals. We evaluate SCO on GPS velocity and electroencephalogram (EEG) data to investigate the optimized curve, parameters, derivatives, and integrals of the final fit. The complete open-source project implementation and technical documentation are available on GitHub.

## II. Methodology

Let the raw signal be represented as two discrete sequences of equal length *N*, denoted as $x = [x_1, x_2, \dots, x_N]^T$ and $y = [y_1, y_2, \dots, y_N]^T$, where $x$ is the independent variable, and $y$ denotes the corresponding observations. The original signal is partitioned into $K$ segments. Let the set of segment *x*-value boundaries be defined as $\psi = [\psi_1, \psi_2, \dots, \psi_{K+1}]^T$, where $\psi_1$ is $x_1$ and $\psi_{K+1}$ is $x_N$, such that each *k*-th segment $k = 1,2,\dots,K$ spans the interval $[\psi_k, \psi_{k+1}]$. Furthermore, segment boundary indices are represented as $D = [d_1, d_2, \dots, d_{K+1}]^T$, where $d_1$ is 1 and $d_{K+1}$ is *N*. A general user-defined function is denoted as $f(x, \beta)$, which is assumed to be differentiable and numerically stable. For each segment $k$ and parameter vector $\beta_k$, which is subject to user-defined bounds $[L_k, U_k]$, with locally translated *x*-values, the goal of the optimization is to minimize the error between the original signal and the estimating function:

$$\beta_k^* = \arg\min_{\beta_k} \sum_{i=d_k}^{d_{k+1}} [y_i - f(x_i - \psi_k; \beta_k)]^2 \tag{1}$$

### A. Signal preprocessing

Data preprocessing is the initial step to enhance optimization stability and speed. The original *x* and *y* datasets are normalized with standard scaling using their mean and standard deviation:

$$\tilde{x} = \frac{x - \mu_x}{\sigma_x}, \quad \tilde{y} = \frac{y - \mu_y}{\sigma_y} \tag{2}$$

Where $\sigma_x, \sigma_y$ represent the standard deviations and $\mu_x, \mu_y$ denote the means for the original datasets. Segment boundary *x*-values are scaled as:

$$\tilde{\psi} = \frac{\psi - \mu_x}{\sigma_x} \tag{3}$$

Additionally, each segment is locally translated to prevent parameters from being influenced by the global *x* magnitude. This is critical for higher stability and meaningful interpretation of the final parameters. For each segment $k$, $\tilde{\psi}_k$ represents the normalized *x*-value at the boundary, and $\tilde{x}_l$ denotes the corresponding local *x*-values of the segment:

$$\tilde{x}_l = \tilde{x}_i - \tilde{\psi}_k, \quad \text{for } i = d_k, \dots, d_{k+1} \tag{4}$$

## *B. Levenberg-Marquardt Optimization*

Optimization is performed with the Levenberg-Marquardt (LM) algorithm on the high-performance computing library JAX [12]. The main speed advantage of JAX is compiling code straight into optimized machine code via the Accelerated Linear Algebra (XLA) compiler. Initially, LM receives an input of initial guess for each parameter with corresponding lower and upper bounds. The arctanh transformation is used to restrict the solution in user-defined bounds $[L_k, U_k]$. Bounds improve stability and are essential for some models, for example, to limit the frequency of a sine wave based on segment length to prevent overfitting. Original constrained parameters $\beta_k$ are converted to unconstrained space $u_k$ using the arctanh transformation:

$$u_k = \operatorname{arctanh}\left(2 \cdot \frac{\beta_k - L_k}{U_k - L_k} - 1\right) \tag{5}$$

LM optimization is performed in unconstrained space, excluding residuals for which parameters must be converted back to constrained space. Hence, Jacobian matrix $J_k$ is converted to an unconstrained representation $J_u$ by using the chain rule of the arctanh transformation:

$$J_u = J_c \cdot \left(\frac{U_k - L_k}{2} \cdot \operatorname{sech}^2(u_k)\right) \tag{6}$$

The LM algorithm was primarily selected for its fast execution time and high stability for non-linear least-squares problems; the general equation for the optimization is given as:

$$\left({J_u}^T J_u + (\lambda + \alpha)I\right)\Delta u_k = -{J_u}^T r \tag{7}$$

The equation is solved for $\Delta u_k$, where $J_u$ is the Jacobian matrix of the residuals $r$ with respect to the parameters $u$, $\lambda$ is the damping factor, and $\alpha$ is a ridge regularization. The proposed step for each parameter is clipped by $\Delta u_{max}$, and the total vector norm $\|\Delta u_k\|$ is scaled to not exceed $\Delta u_{norm}$. Step clipping does not allow the optimizer to propose large, often inaccurate steps. Lastly, $\alpha$ is scaled with the maximum element in the diagonal of Hessian $({J_u}^T J_u)$ to keep the optimization stable.

## *C. Continuity Enforcement*

Many real-world signals are continuous in time; therefore, to analyze them properly, the final curve must also be continuous. The proposed algorithm enforces $C^0$ (value) and $C^1$ (derivative) continuity at the transitions between adjacent segments. For each segment $k > 1$, two parameters $\gamma_0, \gamma_1 \in \beta_k$, are completely excluded from the LM optimization. Instead, these parameters are derived from the continuity equations during the calculation of residuals:

$$f(0, \beta_k) = f(\tilde{\psi}_k - \tilde{\psi}_{k-1}, \beta_{k-1}) \tag{8}$$

$$f'(0, \beta_k) = f'(\tilde{\psi}_k - \tilde{\psi}_{k-1}, \beta_{k-1}) \tag{9}$$

Continuity equations ensure equal value and slope at the transition between the current segment $k$ and the preceding segment $k-1$. The value parameter $\gamma_0$ is solved from the first equation and the derivative parameter $\gamma_1$ from the second equation. Specifically, $C^1$ continuity can be enforced only if the model possesses a linear term $(\gamma_1 x + \gamma_0)$; otherwise, if only an offset parameter $\gamma_0$ is present, $C^0$ continuity can be achieved. The framework includes presets for fitting with linear, polynomial, trigonometric, Gaussian, logistic, exponential, rational, and Fourier wave models with the initial guesses.

## *D. Stability problems*

Imposing such strict constraints introduces significant non-convexity into the error surface, especially in non-linear models. For example, if a segment begins with unfavorable continuity parameters from the previous segment, optimization becomes highly unstable, and the error propagates forward. Fig. 1 illustrates an example of SCO on a cubic model without batched optimization and forward fit techniques. Hourly Bitcoin data from 1st January 2021 to 11th February 2021 was retrieved for this experiment. The red curve represents the optimized fit, blue points indicate the raw signal, and dashed gray lines show segment boundaries.

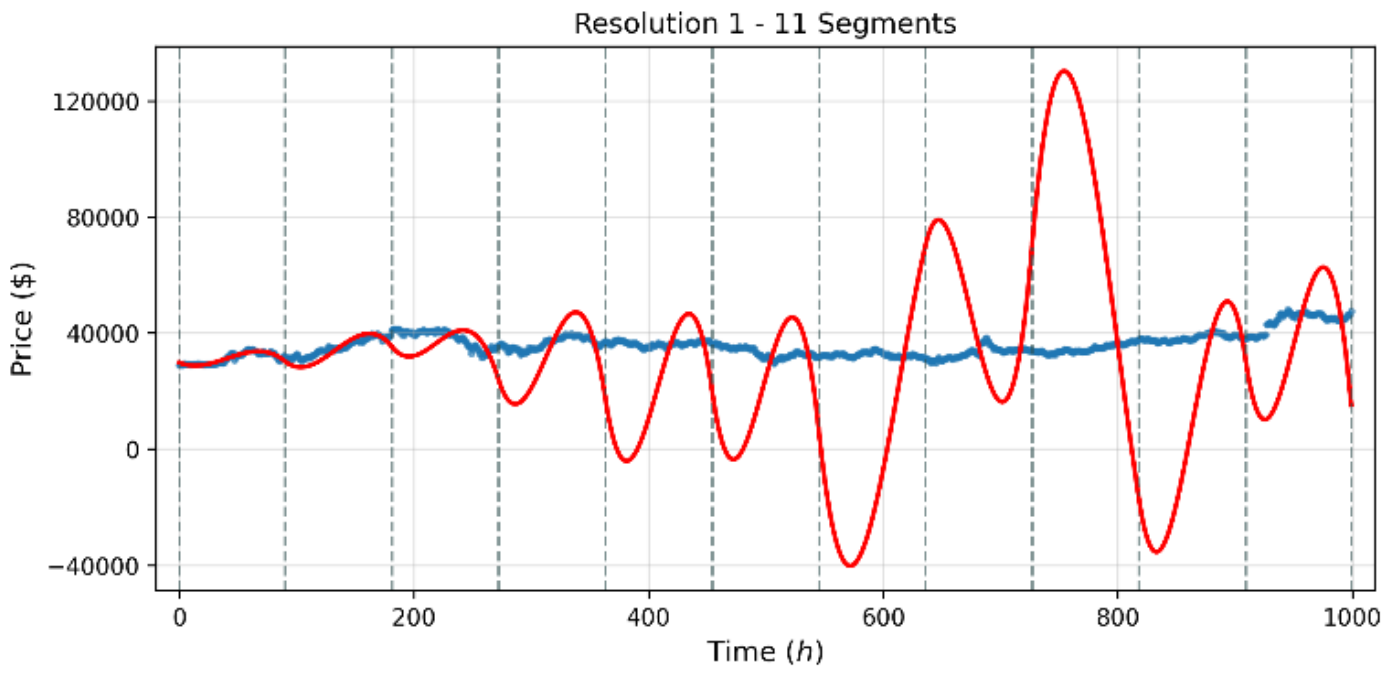

Fig. 1. Continuous piecewise fitting without stability techniques

Initially, SCO accurately followed the raw data; however, after one irrelevant continuity constraint at the third boundary, optimization became unstable with catastrophic error propagation in the subsequent segments. The next section introduces stability techniques to address this limitation.

## *E. Stability Techniques*

Optimization is performed on batches of $q$ segments ($q = 5$ by default) to improve accuracy and stability. For each batch $b$, parameter vectors are stacked together in the form of $\beta_b = [\beta_1, \beta_2, \ldots, \beta_q]^T$ with the stacked bounds $[L_b, U_b]$. If the segment begins with unfavorable continuity constraints from the previous segment, batched optimization can alter the initial segment to find a compromised solution for higher accuracy.

However, relying only on batched optimization is still insufficient for consistently high stability. At the boundary transition between two batches, adverse continuity constraints from the previous batch can still lead to instability. We propose a Forward Fit mechanism to mitigate such error propagation: $q+1$ segments are optimized within each batch, but the last segment is discarded from the fit; instead, it is assigned as an initial guess for the next batch's first segment. The last segment is discarded specifically to re-optimize it in the next batch while already having relevant continuity constraints. Thus, during the optimization of a single batch, LM enforces continuity from the past and ensures that the overall fit is compatible with the future batch, which efficiently solves frequent instability problems.

### *F. Parameter Unscaling*

The optimized parameters must be properly unscaled to retain physical meaning and units. We derive the unscaling formulas by substituting scaling equations for $x$ and $y$ back into the general function $y = f(x, \beta)$. For each segment $k$, let subscript $u$ denote unscaled parameters.

$$\tilde{y} = \frac{y - \mu_y}{\sigma_y} \tag{10}$$

Locally translated $x$-values are indicated as $\tilde{x}_l$ and $\psi_k$ represents the absolute $x$-value at the segment boundary:

$$\tilde{x}_l = \frac{x - \mu_x}{\sigma_x} - \frac{\psi_k - \mu_x}{\sigma_x} = \frac{x - \psi_k}{\sigma_x} \tag{11}$$

For example, on a 5-parameter sine wave ($\sin_5$) defined as:

$$\tilde{y} = A_0 \sin(B_0 \tilde{x}_l + D) + C_1 \tilde{x}_l + C_0 \tag{12}$$

We substitute the scaling equations for $\tilde{y}$ and $\tilde{x}_l$:

$$\frac{y - \mu_y}{\sigma_y} = A_0 \sin\left(B_0 \frac{x - \psi_k}{\sigma_x} + D\right) + C_1 \frac{x - \psi_k}{\sigma_x} + C_0 \tag{13}$$

The whole equation is simplified to find the equations for unscaling parameters. Term $(x - \psi_k)$ is substituted as $x_l$:

$$y = A_0 \sigma_y \sin\left(\frac{B_0}{\sigma_x} x_l + D\right) + \frac{C_1 \sigma_y}{\sigma_x} x_l + C_0 \sigma_y + \mu_y \tag{14}$$

Thus, unscaling equations for each parameter are derived as:

$$A_{0,u} = A_0 \sigma_y \quad B_{0,u} = \frac{B_0}{\sigma_x} \tag{15}$$

$$C_{1,u} = \frac{C_1 \sigma_y}{\sigma_x} \quad C_{0,u} = C_0 \sigma_y + \mu_y \quad D_u = D \tag{16}$$

Finally, Algorithm 1 summarizes the workflow of the SCO algorithm during a single batch.

**Algorithm 1:** Segmented Continuous Optimization

**Input:** original signal $(x, y)$, general function $f(x, \beta)$, changepoints array $(\psi)$, initial guesses $(p_0)$, lower and upper bounds $[L, U]$

1. Input signal and changepoints are normalized using Z-score transformation, and each segment's $x$-values are locally translated.
2. Two specified parameters $\gamma_0, \gamma_1$ from the full vector $\beta_k$ are excluded to be derived later during the calculation of residuals.
3. Initial parameters and bounds are merged into batches of $q + 1$ segments for batched optimization and forward fitting.
4. Residuals are calculated, where $C^0$ and $C^1$ continuity is enforced by solving for $\gamma_0, \gamma_1$ parameters.
5. Parameters and Jacobian matrix are converted to unconstrained space; LM step is calculated and clipped.
6. If the proposed error is less than the previous error, the step is accepted and $\lambda$ is decreased; otherwise, the step is rejected, $\lambda$ is increased, and parameters remain unchanged.
7. Repeat steps from 4 to 6 until converged or maximum iterations reached.
8. Optimized parameters are converted to constrained space and the last segment of the batch is excluded from the fit and assigned as the initial guess for the next batch's first segment.

## III. Experiments And Tests

In this section, we evaluate the accuracy and speed of the proposed SCO algorithm. Tests were performed on non-stationary time-series data, including Bitcoin, velocity, and EEG signals, to demonstrate its efficiency and practical usage. Experiments were executed on a computer with Intel i7-12700 CPU, 16 GB RAM, and NVIDIA RTX 3050 GPU, running on Windows 11 64-bit, Python 3.9.

### *A. Algorithm Demonstration*

SCO was tested on non-stationary daily Bitcoin data from January 2021 to March 2023. Fig. 2 shows the output of the SCO algorithm using the 5-parameter sinusoidal model presented in (12) with uniform segmentation; the legend is consistent with Fig. 1. We use 6 resolutions to reveal how SCO describes a signal in different contexts of short-term and long-term patterns. For instance, optimized parameters for the first segment of resolution 6 for $x \in [0,10]$ are presented as follows:

$$f(x) = 12072 \sin(0.041x - 2.12) + 19.85x + 46038 \tag{17}$$

Final parameters allow us to estimate local and global frequency ($b_0$), volatility ($A$), and trend ($c_1$). Overall, SCO maintained a local Standardized Root Mean Square Error (SRMSE) of 0.553 with an execution time of 2.01 seconds.

### *B. Accuracy and Speed Tests*

The proposed algorithm was evaluated on 30 datasets of different volatility, structure, and scales on all default models. Test datasets include 6 major cryptocurrency markets [13] on a second, minute, hourly, and daily time intervals on different scales, with an additional 6 datasets of extreme scales and zero volatility. For each segment $k$ and its size $N_k = d_{k+1} - d_k + 1$, segment $y$-values are represented as $y_i$ and predicted values as $\hat{y}_i$. The primary metric for the measurement of accuracy is Root Mean Square Error (RMSE) scaled by the standard deviation of the segment $\sigma_k$ (local SRMSE), calculated as follows:

$$\mathrm{SRMSE}_k = \frac{\sqrt{\frac{1}{N_k} \sum_{i=d_k}^{d_{k+1}} (y_i - \hat{y}_i)^2}}{\sigma_k} \tag{18}$$

Multi-scale SCO is utilized on all included models with $M$ resolutions to evaluate accuracy across a different number of segments (as in Fig. 2). Because of the absence of the $\gamma_1 x$ term for $C^1$ continuity, logistic, exponential, rational, and Gaussian functions achieve only $C^0$ continuity.

The local SRMSE metric can become unstable on nearly flat segments, yielding high errors that inflate the accuracy measure. Therefore, segments with a deviation of less than 1% of the $y$-dataset deviation are considered flat, and their local SRMSE is set to 1.0, marking them as a neutral fit. The local SRMSE refers to the average SRMSE for each segment independently, where global SRMSE is calculated for the whole dataset using $\sigma_y$. Global SRMSE is split into two metrics: high $K$, which indicates the average error for the first half of resolutions, and low $K$ for the second half of resolutions. Results are summarized in Table I with standard deviations of each metric to evaluate stability.

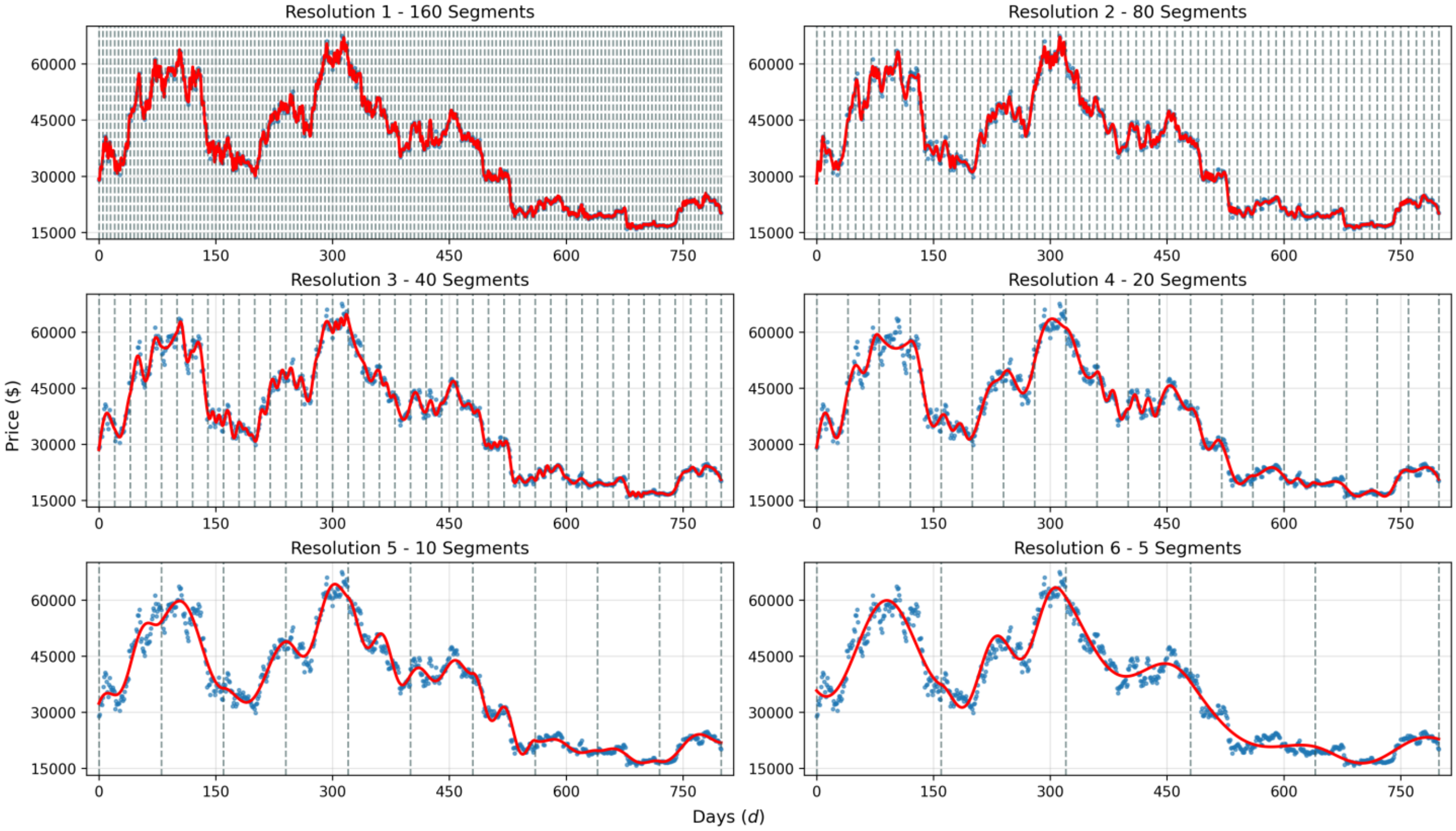

Fig. 2. Demonstration of SCO on Bitcoin daily data

TABLE I. ACCURACY OF THE SCO ALGORITHM

| Model Type | Model | Local SRMSE | Global SRMSE (high *K*) | Global SRMSE (low *K*) |
|---|---|---|---|---|
| Cubic | $ax^3 + bx^2 + cx + d$ | 0.729 ± 0.294 | 0.042 ± 0.013 | 0.173 ± 0.079 |
| $\text{Sin}_5$ | $A_0 \sin(B_0 x + D) + C_1 x + C_0$ | 0.673 ± 0.316 | 0.038 ± 0.012 | 0.160 ± 0.073 |
| Logistic | $\frac{L}{1 + e^{-k(x-x_0)}} + c$ | 0.681 ± 0.324 | 0.039 ± 0.013 | 0.164 ± 0.072 |
| Decay | $a \cdot e^{-bx} + c$ | 0.864 ± 0.334 | 0.055 ± 0.017 | 0.230 ± 0.101 |
| Rational | $\frac{ax}{cx + 1} + b$ | 0.813 ± 0.292 | 0.049 ± 0.016 | 0.210 ± 0.095 |
| Gaussian | $A \cdot e^{-\frac{(x-x_0)^2}{2\sigma^2}} + c$ | 0.683 ± 0.353 | 0.039 ± 0.013 | 0.160 ± 0.072 |
| Fourier | $\sum_{i=1}^{3} a_i \sin(ifx + \phi_i) + (c_1 x + c_0)$ | 0.911 ± 1.897 | 0.078 ± 0.042 | 0.139 ± 0.061 |

These results show a detailed comparison of accuracy and stability across different models. An SRMSE < 1.0 indicates that the fit error is lower than the data's inherent volatility. For local and global SRMSE, $\text{sin}_5$, Gaussian, and logistic models achieved the most accurate results, mainly because of higher complexity and flexibility. Conversely, simpler functions, such as cubic, decay, and rational yield higher error; average stability is roughly equivalent for all models. The Fourier model is an exception, it is naturally unstable and highly complex, which causes the optimization to fail on short and flat segments. However, it still achieves the lowest error for representing global patterns. Overall, higher segmentation mainly results in overfitting, whereas global resolutions result in higher error.

SCO was designed and optimized for high-speed processing. Computational efficiency was tested on all included models with different lengths of input signal *N* and number of segments *K*. For each model, speed is shown in the form of 'first run/subsequent runs' to account for the overhead of JAX Just-in-Time compilation. Results are shown in Table II.

TABLE II. EFFICIENCY OF THE SCO ALGORITHM

| Model Type | $N=10^2$ $K=10^1$ | $N=10^3$ $K=10^2$ | $N=10^4$ $K=10^3$ | $N=10^5$ $K=10^4$ |
|---|---|---|---|---|
| Cubic | 1.36 / 0.0088 | 1.26 / 0.069 | 1.91 / 0.59 | 7.28 / 6.18 |
| $Sin_5$ | 1.55 / 0.0077 | 1.38 / 0.055 | 1.83 / 0.54 | 6.77 / 5.48 |
| Logistic | 1.44 / 0.0044 | 1.28 / 0.034 | 1.51 / 0.32 | 4.49 / 3.29 |
| Decay | 1.30 / 0.0047 | 1.07 / 0.030 | 1.39 / 0.28 | 3.91 / 2.84 |
| Rational | 1.43 / 0.0040 | 1.10 / 0.024 | 1.37 / 0.23 | 3.43 / 2.31 |
| Gaussian | 1.54 / 0.0049 | 1.29 / 0.029 | 1.47 / 0.26 | 3.98 / 2.68 |
| Fourier | 2.03 / 0.019 | 1.81 / 0.086 | 2.64 / 0.88 | 10.65 / 8.93 |

As expected, the first run is generally slower due to the initial JAX Just-In-Time (JIT) compilation. The framework can be compiled initially by using the warmup configuration; subsequent runs show the actual speed of the SCO algorithm without compilation time. Rational, Gaussian, and decay models demonstrated the highest speed, where complex models require more time for optimization. Optimization speed and compilation primarily depend on model complexity, $N$, $K$, quality of initial guesses and bounds.

## C. Velocity Experiment

To evaluate the SCO algorithm further, we used the UAH-DriveSet dataset [14]. The UAH-DriveSet is a public collection of data captured by DriveSafe driving monitoring app by various testers in different environments. Specifically, a 1-lane secondary route with normal behavior was chosen to retrieve GPS velocity data. A cubic model with 12 segments was used to depict the local patterns in velocity and derive analytical acceleration from the optimized fit shown in Fig. 3.

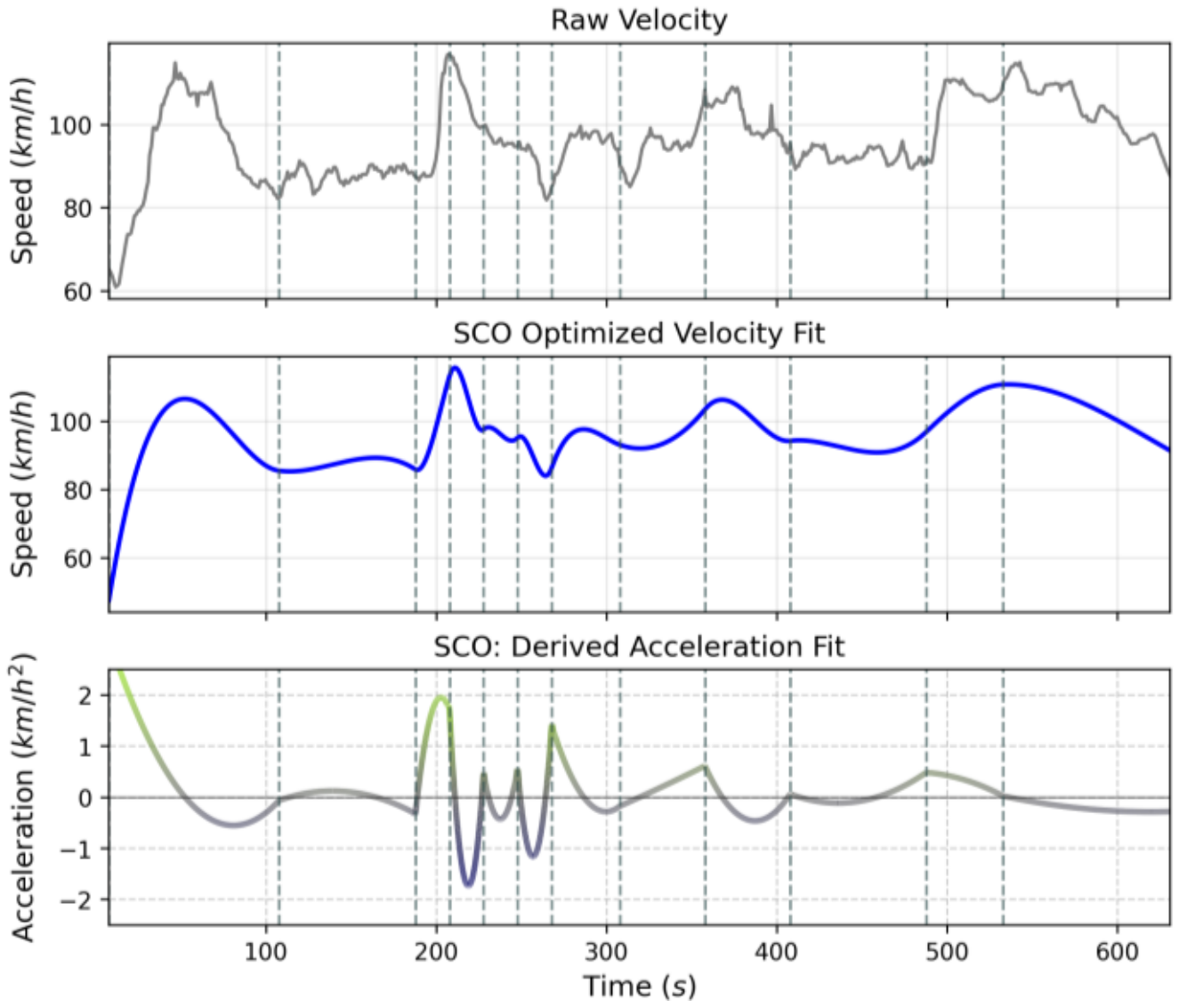

Fig. 3. Demonstration of SCO on velocity data and derived acceleration

As observed, the SCO framework accurately follows raw velocity and acceleration, extracting optimized parameters to examine underlying signal dynamics. Non-uniform segmentation was used based on the change in signal volatility for higher accuracy and interpretability. Because the velocity fit is $C^1$ continuous, the resulting acceleration fit remains $C^0$ continuous. Throughout the experiment, we observed that the absolute value of the acceleration depends on the length of the segments. This is caused by local polynomial coefficients, which decrease when the magnitude of $x$ increases in longer segments, resulting in a lower scale of the derivative. Therefore, to address this, the acceleration graph is visualized using normalized colors with dark blue for negative and yellow for positive maximum values. Overall, SCO achieved an average local SRMSE of 0.629 in 1.84 seconds.

## D. EEG Analysis

Our algorithm was applied to describe electroencephalogram (EEG) signals [15,16] in 13 segments using a 5-parameter sine model shown in Fig. 4. The dataset contains EEG recordings of subjects during the performance of a mental arithmetic task. The arithmetic task was the serial subtraction of two numbers. Each trial started with the communication of orally 4-digit (minuend) and 2-digit (subtrahend) numbers (e.g. 3141 and 42). Additionally, the integral of the optimized curve was calculated to further examine the signal. To retain physical plausibility, non-negative bounds for amplitude and frequency were used.

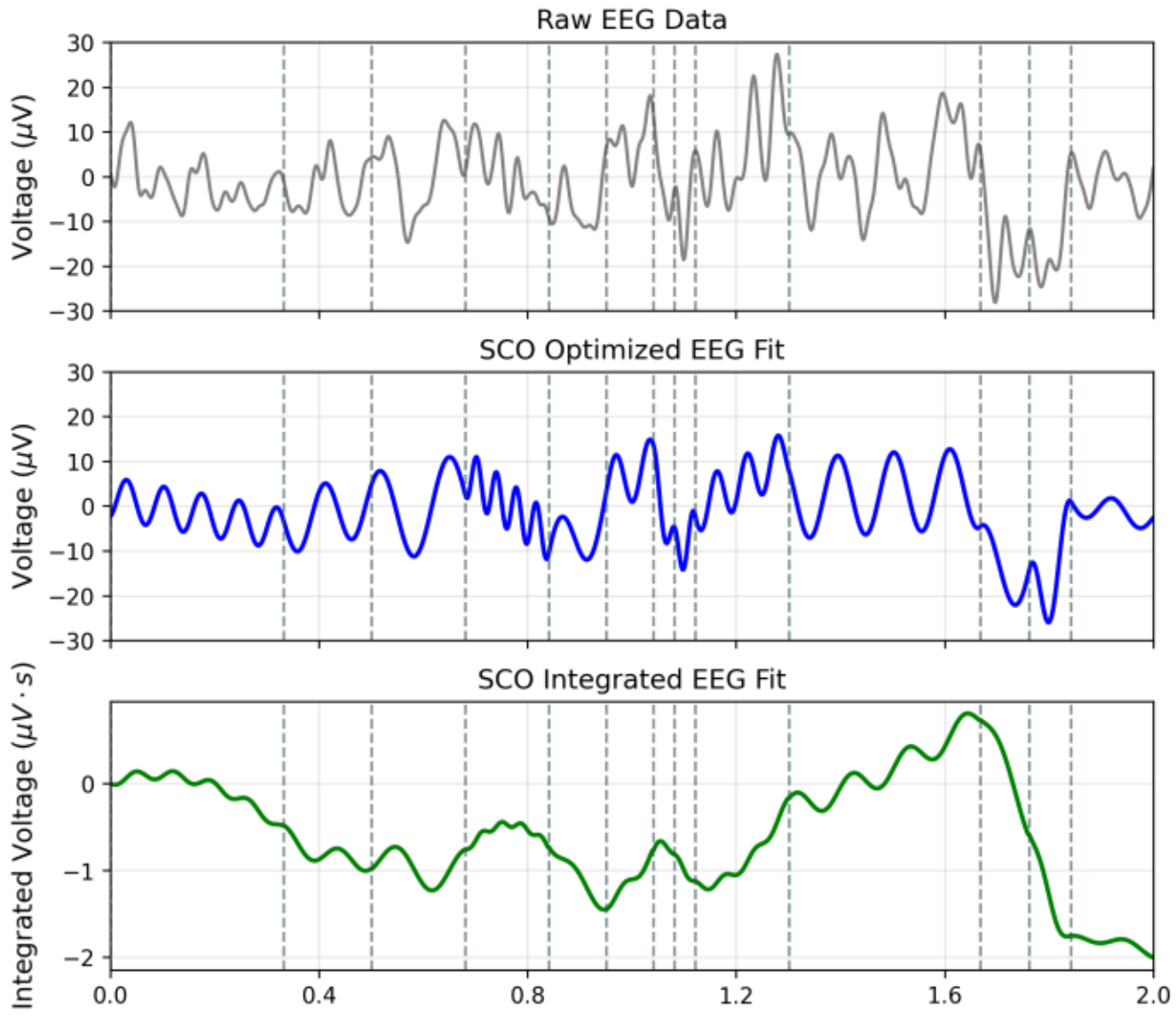

Fig. 4. SCO experiment on the EEG signal and integrated voltage

Optimized parameters are used to analyze the inherent properties of the signal's local and global frequency, amplitude and trend. Moreover, the EEG signal was analytically integrated to represent the total accumulated voltage during the performance of the arithmetical task, which can be used for a more comprehensive investigation. SCO maintained an average local SRMSE of 0.552 and execution time of 1.59 seconds.

## IV. Conclusion

This paper presented the Segmented Continuous Optimization framework, a novel algorithm for representing non-stationary signals with a user-defined function. SCO addresses the limitations of traditional regression algorithms by performing segmented, continuous curve fitting to appropriately analyze local and global patterns in the data. The proposed algorithm was thoroughly tested on non-stationary datasets of different structure, scale, and volatility. Results demonstrate the practical usage of the SCO algorithm with applications across various fields. Future work will focus on real-time implementation with improved speed and accuracy, the derivation of additional parameters for higher levels of continuity, and a wider range of included presets.

## References

[1] K. Ulm, "A statistical method for assessing a threshold in epidemiological studies," *Statistics in Medicine*, vol. 10, no. 3, pp. 341–349, 1991, doi: 10.1002/sim.4780100306

[2] F. Zhang and Q. Li, "Robust bent line regression," *Journal of Statistical Planning and Inference*, vol. 185, pp. 41–55, 2017, doi: 10.1016/j.jspi.2017.01.001

[3] J. Liu, S. Wu, and J. V. Zidek, "On segmented multivariate regression," *Statistica Sinica*, vol. 7, no. 3, pp. 497–525, 1997, url: http://www.jstor.org/stable/24306090

[4] V. M. R. Muggeo, "Estimating Regression Models with Unknown Break-Points," *Statistics in Medicine*, vol. 22, no. 19, pp. 3055–3071, October 2003, doi: 10.1002/sim.1545

[5] J. H. Friedman, "Multivariate Adaptive Regression Splines," *The Annals of Statistics*, vol. 19, no. 1, pp. 1–67, March 1991, url: https://www.jstor.org/stable/2241837

[6] J. Duan, Q. Wang and Y. -P. Wang, "HOPS: A Fast Algorithm for Segmenting Piecewise Polynomials of Arbitrary Orders," in *IEEE Access*, vol. 9, pp. 155977-155987, 2021, doi: 10.1109/ACCESS.2021.3128902.

[7] R. Tibshirani, "Adaptive piecewise polynomial estimation via trend filtering," *The Annals of Statistics*, vol. 41, no. 6, pp. 2857–2900, December 2013, doi: 10.1214/13-AOS1189.

[8] M. Unser, "Splines: a perfect fit for signal and image processing," in IEEE Signal Processing Magazine, vol. 16, no. 6, pp. 22-38, Nov. 1999, doi: 10.1109/79.799930.

[9] K.-P. Lu and S.-T. Chang, "An Advanced Segmentation Approach to Piecewise Regression Models," *Mathematics*, vol. 11, no. 24, p. 4959, December 2023, doi: 10.3390/math11244959.

[10] K. Levenberg, "A method for the solution of certain non-linear problems in least squares," *Quarterly of Applied Mathematics*, vol. 2, no. 2, pp. 164–168, 1944, doi: 10.1090/qam/10666.

[11] D. W. Marquardt, "An Algorithm for Least-Squares Estimation of Nonlinear Parameters," *Journal of the Society for Industrial and Applied Mathematics*, vol. 11, no. 2, pp. 431–441, 1963, doi: 10.1137/0111030

[12] J. Bradbury *et al.*, "JAX: composable transformations of Python+NumPy programs," version 0.3.13, 2018. [Online]. Available: http://github.com/jax-ml/jax

[13] "CCXT – Crypto Currency eXchange Trading Library," version 4.4.0, 2026. [Online]. Available: https://github.com/ccxt/ccxt

[14] E. Romera, L. M. Bergasa, and R. Arroyo, "Need Data for Driving Behavior Analysis? Presenting the Public UAH-DriveSet," in *IEEE International Conference on Intelligent Transportation Systems (ITSC)*, Rio de Janeiro, Brazil, Nov. 2016, pp. 387–392, doi: 10.1109/ITSC.2016.7795584.

[15] A. Goldberger, L. Amaral, L. Glass, J. Hausdorff, P. C. Ivanov, R. Mark, *et al.*, "PhysioBank, PhysioToolkit, and PhysioNet: Components of a new research resource for complex physiologic signals," *Circulation*, vol. 101, no. 23, pp. e215–e220, June 2000. [Online]. Available: https://www.ahajournals.org/doi/10.1161/01.CIR.101.23.e215

[16] I. Zyma, S. Tukaev, I. Seleznov, K. Kiyono, A. Popov, M. Chernykh, and O. Shpenkov, "Electroencephalograms during Mental Arithmetic Task Performance," Data, vol. 4, no. 1, p. 14, Mar. 2019, doi: 10.3390/data4010014.